\documentclass[final,1p,times]{elsarticle} 

\usepackage{graphicx}
\usepackage{amssymb} 
\usepackage{amsthm} 
\usepackage{lineno}

\def\SCETG{{\rm SCET}_{\rm G}\,}
\newcommand{\mcdot}{\!\cdot\!}

\newcommand{\e}{\mathrm{e}}

\newcommand{\be}{\begin{equation}}
\newcommand{\ee}{\end{equation}}

\def\OMIT#1{{}}

\newcommand{\vc}[1]{{\bf{#1}}}

\def\bnslash{{\overline{n}}\!\!\!\slash}

\journal{Nuclear Physics A} 

\begin{document}

\begin{frontmatter} 

\title{Medium-induced splitting kernels from $\SCETG$}

\author{Grigory Ovanesyan}
\address{Los Alamos National Laboratory, Theoretical Division, Mail Stop B283,  Los Alamos, NM 87544, U.S.A.}


\begin{abstract} 
Using the framework of soft-collinear effective theory with Glauber gluons ($\SCETG$), we evaluate medium-induced splitting kernels. Because of the power counting of the effective theory, our results are valid for arbitrary, not necessarily small values of the energy fraction $x$ taken by the emitted parton. In this framework we prove the factorization from the hard process and gauge invariance of the splitting kernels, we also show how nuclear recoil and the phase space cuts can be implemented into the phenomenology.
\end{abstract} 

\end{frontmatter} 


\section{Introduction}
Jet quenching phenomenology based on perturbative QCD and energy loss calculations has been quite successful in describing data from RHIC and LHC heavy ion collisions, see \cite{Gyulassy:2003mc,CMS:2012aa} and the references therein. Typically, such calculations incorporate scattering of highly energetic quarks and gluons produced in the medium off of the QGP quasiparticles. The resulting energy loss and the steepness of the production cross section of quarks and gluons combine together and lead to the phenomenon of jet quenching.

In order to go beyond the approximations used in energy loss calculations we take the effective theory point of view, without assuming that the emitted partons are soft. A good start for an effective theory for jet propagation in the medium is soft-collinear effective theory (SCET) \cite{Bauer:2000ew,Bauer:2000yr,Bauer:2001ct,Bauer:2001yt}. In this effective theory all hard modes are integrated out and only quark and gluons with momenta collinear to the directions of interest and ultrasoft gluons are present. In order to adapt SCET for proper description of jets in the medium, the elastic scattering of energetic quarks and gluons off of the medium sources has to be included. This elastic scattering is mediated by transverse $t-$channel gluons, which are traditionally called Glauber gluons. Inclusion of this mode leads to an effective theory, which we call $\SCETG$, which stands for soft collinear effective theory with Glauber gluons. This effective theory has been developed in a series of papers \cite{Idilbi:2008vm,D'Eramo:2010ak,Ovanesyan:2011xy}. Also, note that it has been shown in Ref. \cite{Bauer:2010cc} that in order to consistently describe Drell-Yan process in effective theory one needs to go beyond SCET and use in fact, $\SCETG$ and include the potential scattering of spectators off each other.

In this talk we overview main results obtained in Ref. \cite{Ovanesyan:2011kn} for medium-induced splitting kernels. In the framework of $\SCETG$ we present all  splitting kernels keeping the full emitted parton energy fraction $x$ dependence, which is consistent with the power counting of our effective theory.  

\section{Effective theory for jets in the dense QCD medium}
The momentum of the $t-$channel gluon mode that mediates the elastic scattering of jets off of the QGP quasiparticles can vary from $q\sim(\lambda^2, \lambda^2, \lambda)$ for static source to $q\sim(\lambda, \lambda^2,\lambda)$  for soft source \cite{Ovanesyan:2011xy}. We refer to gluons with momentum scaling of both of these modes as Glauber gluons, even though traditionally under this name the former mode is understood. Both modes are off-shell and from effective theory point of view the role of these modes is to provide interaction with the potential created by the source. Thus, the appropriate treatment for such interaction would be a term in the Lagrangian coupled with background vector potential. This has been done in Ref. \cite{Idilbi:2008vm} for a collinear quark field and extended in Ref. \cite{Ovanesyan:2011xy} for a collinear gluon field interacting with external vector potential. The details of this extra term in the Lagrangian that needs to be added to SCET  depend on the gauge fixing condition and the underlying assumption on the momentum scaling of the source field that produces the elastic scattering. Different gauges and different choices of scaling for the source have been considered in Ref. \cite{Ovanesyan:2011xy}. Assuming static quark center with momentum scaling of HQET field, i.e. $p\sim(1,1,\lambda)$ for the source and a particularly simple gauge choice, where Glauber gluons are quantized in the covariant gauge and collinear gluons are quantized in the light-cone gauge, which we call the hybrid gauge, we get the following effective Lagrangian of $\SCETG$\cite{Ovanesyan:2011xy}:
\begin{eqnarray}
&&{\mathcal{L}}_{\SCETG}(\xi_n, A_n, A_{G})=\mathcal{L}_{\rm{SCET}}(\xi_n, A_n)+\nonumber\\
&&\qquad\qquad\qquad\qquad g\sum_{p,p'}\,\e^{-i(p-p')x}\left(\overline{\xi}_{n,p'}T^a\frac{\bnslash}{2}\xi_{n,p}-if^{abc}A^{\lambda c}_{n,p'}A^{\nu b }_{n,p}\,g_{\nu\lambda}^{\perp}\,\overline{n}\mcdot p\right)\,n\mcdot A^a_{G}(x).
\end{eqnarray}
The first term in the second line describes elastic scattering of a collinear quark field off of the scattering center and the second term describes elastic scattering of a collinear gluon field off of the scattering center in the medium. Note that different collinear sectors with momenta $p$ and $p'$ separated in the transverse direction are coupled through this term.
\begin{figure}[!t]
\centerline{
\includegraphics[width = 0.65\linewidth]{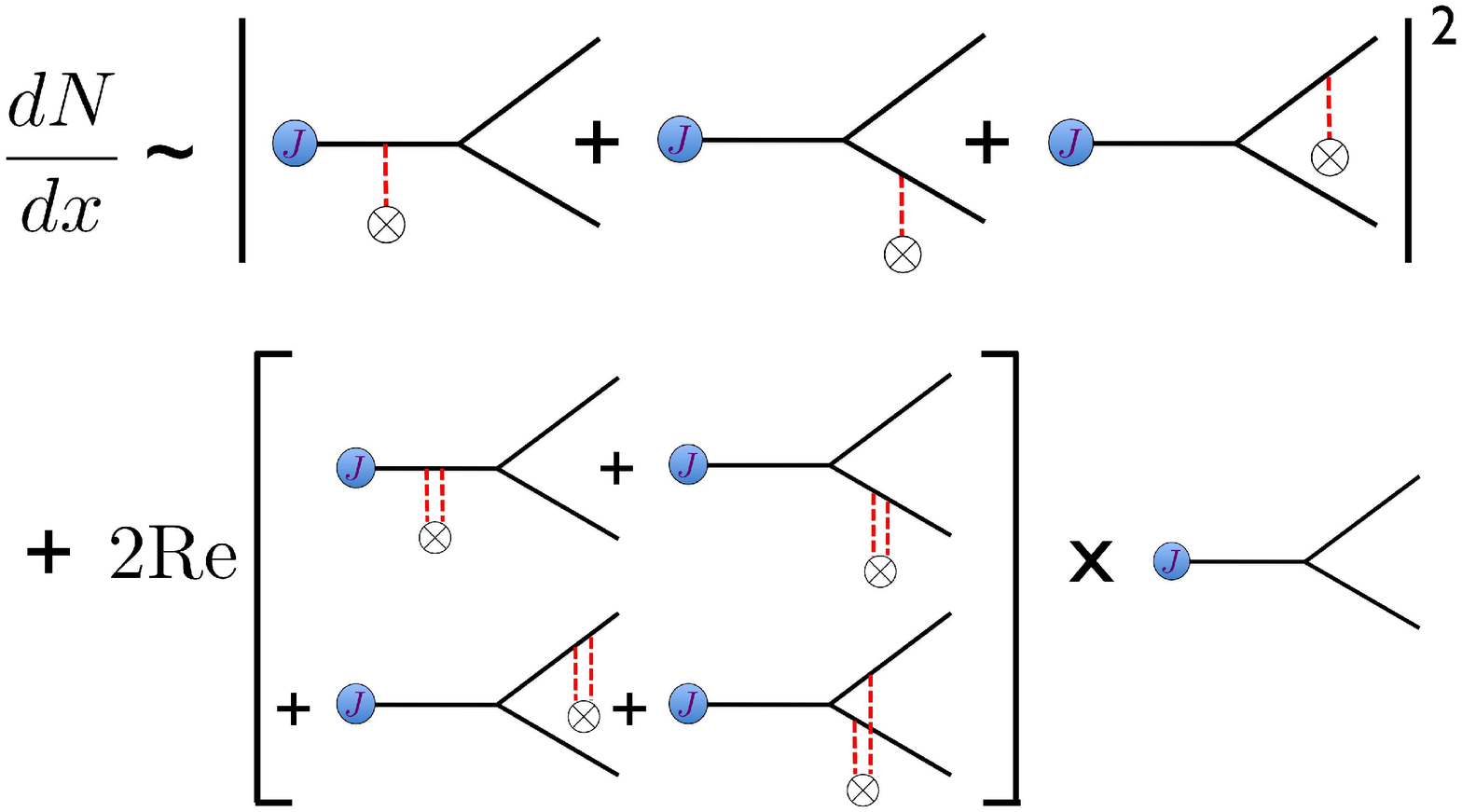}
}
\caption{Single and double Born Feynman diagrams contributing to medium-induced splitting kernels at first order in opacity.
}
\label{fig1}
\end{figure}
\section{Medium-induced splitting kernels}
With the effective Lagrangian of $\SCETG$ at hand, it is a straightforward task to calculate the medium-induced splitting kernels. We consider an initial quark or gluon with momentum $p_0$ splitting into two partons with momenta $p$ and $k$, where the splitting energy fraction is defined as $x=\overline{n}\mcdot k/\overline{n}\mcdot p_0$. We use the following momentum assignment to partons in the splitting: $p_0\rightarrow pk$, i.e. for example when we refer to $q\rightarrow qg$ splitting we have in mind that a quark with momentum $p_0$ splits into a  quark with momentum $p$ and a gluon with momentum $k$. We thus have four possible splittings: $q\rightarrow qg$, $g\rightarrow gg$, $g\rightarrow q\overline{q}$ and $q\rightarrow gq$. The final partons are taken to be on-shell: $k^2=p^2=0$. The light-cone vector $n$ is aligned with the initial parton, so that by definition $\vc{p}_{0\perp}=0$ and thus $\vc{k}_{\perp}=-\vc{p}_{\perp}$. The diagrams in the medium that need to be evaluated are presented in Figure \ref{fig1}. The result of applying the Feynman rules of $\SCETG$ and evaluating single and double Born diagrams along with longitudinal integrals can be expressed through four transverse momenta which we define as:
\begin{eqnarray}
\vc{A}_{\perp}=\vc{k}_{\perp}, \qquad \vc{B}_{\perp}=\vc{k}_{\perp}+x\,\vc{q}_{\perp}, \qquad \vc{C}_{\perp}=\vc{k}_{\perp}-(1-x)\,\vc{q}_{\perp},\qquad \vc{D}_{\perp}=\vc{k}_{\perp}-\vc{q}_{\perp}.
\end{eqnarray}
After a straightforward though lengthy calculation we obtain the following result for arbitrary medium-induced splitting\cite{Ovanesyan:2011kn}:
\begin{eqnarray}
&&\frac{{\rm{d}}N^{(i)}}{{\rm{d}}x \,{\rm{d}}^2\vc{k}_{\perp}}=\frac{\alpha_s}{2\pi^2}\,{P^{(i)}_{{\rm{vac}}}(x)}\int \frac{{{\rm{d}}\Delta z}}{\lambda_{i}(z)}\,{\rm{d}}^2\vc{q}_{\perp}\frac{1}{\sigma_{{\rm{el}}}}\frac{{\rm{d}}\sigma_{{\rm{el}}}}{{\rm{d}}^2\vc{q}_{\perp}}\,\sum_{k=1}^5\,\alpha_k^{(i)}(1-\cos\Phi_k),
\end{eqnarray}
where $\sigma_{{\rm{el}}}$ is the elastic scattering cross section of the parton in the medium and  as suggested by notation the phases $\Phi_1-\Phi_5$ are independent of the splittings. They are equal to:
\begin{figure}[!t]
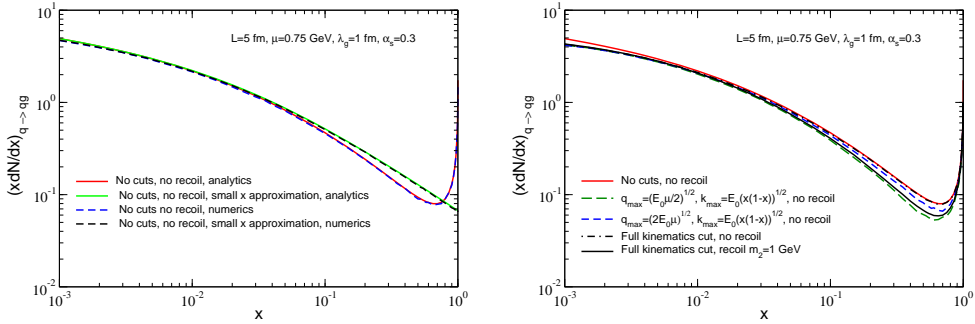

\label{fig2}
\centerline{
\includegraphics[width = 0.45\linewidth]{fig1.eps} \hspace*{4mm}
\includegraphics[width = 0.45\linewidth]{fig5.eps}
}
\caption{Left panel: medium-induced splitting $q\rightarrow qg$ with no nuclear recoil and infinite phase-space cuts. Red and green solid lines correspond to full splitting and soft $x$ approximation respectively, using analytical formulas from Ref. \cite{Ovanesyan:2011kn}, same colors but dashed lines represent numerical integration. Right panel: medium-induced splitting $q\rightarrow qg$ taking into account medium-recoil and finite phase-space cuts. Different lines correspond to: solid red is with infinite cuts and no medium recoil, dashed green is cut on Glauber momenta ${q}_{\perp {\rm{max}}}=\sqrt{E_0\mu/2}$, cut on emitted gluon transverse momentum ${k}_{\perp {\rm{max}}}=E_0\sqrt{x(1-x)}$ and no medium recoil, dashed blue is for cuts ${q}_{\perp {\rm{max}}}=\sqrt{2E_0\mu}$ and ${k}_{\perp {\rm{max}}}=E_0\sqrt{x(1-x)}$ and no recoil, dot-dashed black corresponds to full kinematics cut and no nuclear recoil and finally, solid black corresponds to full kinematics cut and nuclear recoil for source mass $m_{2}=1\,$GeV. Here $E_0$ is the initial quark energy and $\mu$ is the effective gluon mass in the medium.}
\label{fig2}
\end{figure}

\begin{eqnarray}
&&\Phi_1=\Psi\,\vc{B}_{\perp}^2, \quad\Phi_2=\Psi\,\vc{C}_{\perp}^2, \quad \Phi_3=\Psi\left(\vc{C}_{\perp}^2-\vc{B}_{\perp}^2\right),\qquad\Phi_4=\Psi\,\vc{A}_{\perp}^2, \qquad\Phi_5=\Psi\left(\vc{A}_{\perp}^2-\vc{D}_{\perp}^2\right),\nonumber\\
&&{\rm{and}}\,\,\,  \Psi=\frac{\Delta z}{x(1-x)\,\overline{n}\mcdot p_0}.
\end{eqnarray}
The coefficients $\alpha^{(i)}_k$ and the scattering lengths $\lambda_i$ in fact depend on the splitting. The scattering lengths are  equal to the gluon scattering length $\lambda_g$ for splittings $q\rightarrow qg$ and $g\rightarrow gg $ and the quark scattering length $\lambda_q$ for the splittings $g\rightarrow q\overline{q}$ and $q\rightarrow gq$. The coefficients $\alpha_k$ for the $q\rightarrow qg$ splitting are equal to:
\begin{eqnarray}
&&\alpha^{(1)}_1=\frac{\vc{B}_{\perp}}{\vc{B}_{\perp}^2}\mcdot\left(\frac{\vc{B}_{\perp}}{\vc{B}_{\perp}^2}-\frac{\vc{C}_{\perp}}{\vc{C}_{\perp}^2}+\frac{1}{N_c^2}\left(\frac{\vc{A}_{\perp}}{\vc{A}_{\perp}^2}-\frac{\vc{B}_{\perp}}{\vc{B}_{\perp}^2}\right)\right), \qquad\,\,\,\,\,\,\alpha^{(1)}_2=\frac{\vc{C}_{\perp}}{\vc{C}_{\perp}^2}\mcdot\left(2\frac{\vc{C}_{\perp}}{\vc{C}_{\perp}^2}-\frac{\vc{A}_{\perp}}{\vc{A}_{\perp}^2}-\frac{\vc{B}_{\perp}}{\vc{B}_{\perp}^2}\right),\nonumber\\
&&\alpha^{(1)}_3=\frac{\vc{B}_{\perp}}{\vc{B}_{\perp}^2}\mcdot \frac{\vc{C}_{\perp}}{\vc{C}_{\perp}^2},\qquad \alpha^{(1)}_4=\frac{\vc{A}_{\perp}}{\vc{A}_{\perp}^2}\mcdot\left(\frac{\vc{D}_{\perp}}{\vc{D}_{\perp}^2}-\frac{\vc{A}_{\perp}}{\vc{A}_{\perp}^2}\right),\qquad\alpha^{(1)}_5=-\frac{\vc{A}_{\perp}}{\vc{A}_{\perp}^2}\mcdot \frac{\vc{D}_{\perp}}{\vc{D}_{\perp}^2},
\end{eqnarray}
and for three remaining splittings they all have been calculated and can easily be extracted from Ref. \cite{Ovanesyan:2011kn}. The vacuum splitting functions $P^{(i)}_{{\rm{vac}}}(x)$ are well known \cite{Altarelli:1977zs} and we do not present them here for brevity. The numerical evaluation of splitting intensity $xdN/dx$ for the splitting $q\rightarrow qg$ is shown in Figure \ref{fig2}. For details on implementation of nuclear recoil and phase space cuts see Ref.  \cite{Ovanesyan:2011xy,Ovanesyan:2011kn}.

In the process of deriving the medium-induced splitting kernels in Ref.  \cite{Ovanesyan:2011kn} we have shown explicitly that they are  gauge invariant and that they factorize from the hard scattering process that creates the jet. This is described in detail in Ref.  \cite{Ovanesyan:2011xy} for $q\rightarrow qg$ splitting and works similarly for the remaining splittings.

\section{Summary}
Using $\SCETG$, several theoretical advances were achieved, relevant for jet quenching phenomenology. We proved the gauge invariance and factorization of the medium-induced splitting kernels. On the quantitative level we calculated the medium-splitting kernels beyond the small$-x$ approximation which is inherent to energy loss calculations. With these improved results for $q\rightarrow qg$ and $g\rightarrow gg$ splittings, as well as two additional splittings  $g\rightarrow q\overline{q}$ and $q\rightarrow gq$, which only appear at finite $x$ order, we have all ingredients to present improved theoretical predictions for jet production at RHIC and LHC \cite{He:2011pd,Neufeld:2012df}. These results will be presented elsewhere.

\end{document}